\documentclass[12pt]{article}
\usepackage{setspace}
\usepackage{epsfig,amsmath,amssymb} 
\usepackage[all]{xy}
\usepackage{ifpdf} 
\usepackage{graphicx}   % for figures
\input{Qcircuit}
\usepackage{amsfonts}%for tick and cross 
\usepackage{makeidx}  
\usepackage{epigraph}
\usepackage{etoolbox}
\makeatletter
\patchcmd{\epigraph}{\@epitext{#1}}{\itshape\@epitext{#1}}{}{}
\makeatother

\newcommand{\qed}{\hfill \mbox{\raggedright \rule{.07in}{.1in}}}

\renewcommand{\ket}[1]{\left | #1 \right\rangle}
\renewcommand{\bra}[1]{\left \langle #1 \right |}

\newcommand{\qq}[1]{``#1"}

\title{Evolution without evolution, and without ambiguities} 

\author{C. Marletto$^{a}$ and  V. Vedral$^{a,b}$
	\\ {\small $^{a}$ Physics Department, University of Oxford} 
	\\{\small $^{b}$Centre for Quantum Technologies, National University of Singapore}}

\date{October 2016}

\begin{document}

\maketitle

\begin{abstract}

\noindent In quantum theory it is possible to explain time, and dynamics, in terms of {\sl entanglement}. This is the {timeless} approach to time, which assumes that the universe is in a {\sl stationary} state, where two non-interacting subsystems, the \qq{clock} and the \qq{rest}, are entangled. As a consequence, by choosing a suitable observable of the clock, the {\sl relative state} of the rest of the universe evolves unitarily with respect to the variable labelling the clock observable's eigenstates, which is then interpreted as time. This model for an \qq{evolution without evolution} (Page and Wootters, 1983), albeit elegant, has never been developed further, because it was criticised for generating severe ambiguities in the dynamics of the rest of the universe. In this paper we show that there are no such ambiguities; we also update the model, making it amenable to possible new applications.

\end{abstract}

\section{Introduction}

All dynamical laws are affected by a deep problem, \cite{BAR1}. They are formulated in terms of an extrinsic parameter time, which is not itself an element of dynamics and hence it is left unexplained. 

One powerful way of addressing this problem is the \qq{timeless approach} to time. Its logic is elegant: both dynamics and time should emerge from more fundamental elements, chosen so that the dynamics satisfies certain criteria \cite{MITHWE}. For example, when applied to Newtonian physics \cite{BAR1, BAR2}, this approach leads to relational dynamics, where one selects a system as a reference clock, with a particular clock-variable, so as to ensure that  Newton's laws hold when that observable is regarded as time (the so-called \qq{ephemeris time}). This picture, however, still requires motion to be assumed as primitive, thus leaving the appearance of dynamics itself unexplained. 

The same problem as in classical physics arises in quantum theory: time appears as an extrinsic parameter in the equations of motion. In quantum theory there is also a deeper problem - a major obstacle for quantum gravity \cite{ZEH}: time is not a quantum observable, and yet quantum observables depend on it. What precisely is its status, and how can it be reduced to more fundamental elements?

Once again, the timeless approach provides an elegant way out: the Page and Wootters (PW) model, \cite{PAWO}. By analogy with classical physics, that approach aims at selecting a clock and an observable of the clock, so that the Schr\"odinger (or Heisenberg) equation holds on the rest of the universe, with respect to the variable $t$ labelling the eigenvalues of that observable. But since observables in quantum theory are operators, the implementation of this approach turns out to be rather different from its classical counterpart - with some advantages, but also, as we are about to recall, various problems. 

The advantage is that, unlike in the classical scenario, in quantum theory motion does not have to be assumed as primitive: one assumes that the whole universe is in a {\sl stationary state} - i.e., it is an eigenstate of its Hamiltonian. Time and dynamics then emerge in a subsystem of the universe that is entangled with some suitably chosen clock, endowed with an appropriate observable that we shall call {\sl clock observable}. It is important to notice that this is {\sl not} a time operator, but simply an observable (such as, say, a component of the angular momentum) of the system chosen as a clock.

Specifically, by supposing that the Hamiltonian is sufficiently local, it is always possible to regard the universe as consisting of two {\sl non-interacting} subsystems, which we shall call \qq{the clock} and \qq{the rest}. A clock-observable $T$, conjugate to the clock's Hamiltonian, defines a basis of eigenvectors $\ket{t}\;:\;T\ket{t}=t\ket{t}$  (the hands of the clock), where $t$ is a real-valued label. Since $T$ does not commute with the total Hamiltonian of the universe, the overall static state of the universe must  be a {\sl superposition} (or mixture, \cite{VED}) of different eigenstates of the clock observable $T$: as a result, a Schr\"odinger equation can be written for the relative state (in the Everett sense \cite{EVE}) of the rest of the universe (relative to $t$) whose parameter time is nothing but the label $t$ of the states of the clock. (An equivalent construction can be carried out in the Heisenberg picture \cite{PAWO}.) 

As we said, nothing in this construction relies on defining a time operator.Thus, quantum theory provides the means to solve the problem of time via its most profound properties: having pairs of non-commuting observables (in this case, the Hamiltonian of the universe, and the clock observable $T$); and permitting entanglement between subsystems of the universe. Unlike in classical dynamics, there is no need to assume any underlying motion: both time {\sl and} motion are explained in terms of motionless entanglement contained in the state of the universe. 

This elegant model leading to an \qq{evolution without evolution} \cite{PAWO}  has promising features, such as its compatibility with quantum gravity \cite{XXX} and its operational nature, that bodes well for experimental techniques involving quantum clocks \cite{NIST}. Yet, it has never been developed beyond the toy-model stage. This is because it is affected by a number of problems, which, though superficially technical, have been regarded as invalidating the whole approach as a contribution to fundamental physics. For example, Kuchar pointed out problems about the possibility of constructing two-time propagators in this model, \cite{KU} -- these have been thoroughly addressed in \cite{LOY}. There are also conceptual problems, because the model seems to have serious ambiguities that do not arise in relational classical dynamics. Specifically, as pointed out by Albrecht and Iglesias, there seem to be a \qq{clock ambiguity}; there are several, non-equivalent choices of the clock \cite{ALB}, which appear produce an ambiguity in the laws of physics for the rest of the universe: different choices of the clock lead to different Hamiltonians, each corresponding to radically different dynamics in the rest of the universe. So, it would seem that the logic of the timeless approach cannot be applied as directly as in classical physics, because it does not lead to a unique Schr\"odinger equation for the rest of the universe. 

In this paper we show that the clock ambiguities in fact do not arise. To see why they do not arise, one must appeal to the necessary properties for a subsystem to be a good clock -- in particular, that it must be weakly interacting with the rest. We also update the PW model, clarifying what constraints the state of the universe must satisfy in order for the model to be realistic, and how it accommodates an unambiguous notions of the flow of time. As a result of this update, the model becomes applicable to a number of open problems, including potential new applications.

\section{Evolution without evolution}

We shall now review the PW approach, by expressing explicitly what conditions must hold for it to be applicable -- namely:

{\bf Timelessness}. The first condition is that the Universe is \qq{timeless}, i.e., it is in an eigenstate $\ket{\psi}\in \cal{H}$ of its Hamiltonian $H$, which can always be chosen so that

\begin{equation}
H\ket{\psi}=0\;.\label{tot}
\end{equation}

This constraint is compatible with existing approaches to quantum gravity - e.g. the Wheeler-DeWitt equation in a closed universe  \cite{WEDE}, but we regard it as the first of a set of sufficient conditions for a timeless approach to time in quantum theory. Note also that this assumption is compatible with observation, as argued in \cite{PAWO}, because it is impossible empirically to distinguish the situation where \eqref{tot} holds from that where the universe's state is not stationary, because the phases appearing in the state $\ket{\psi}$ are unobservable.

{\bf Good clocks are possible}. The second sufficient condition is that the Hamiltonian includes at least one good clock -- by which we mean a system with a large set of distinguishable states, which interacts only weakly with the rest of the universe; in the ideal case, it should not interact at all. \footnote{That a perfect clock must not interact with anything else is not in contradiction with the fact that for actual clocks synchronisation must occur - indeed the latter, since it requires interactions, is always carried out when the clock is {\sl not} being used as a clock.} So, the Hamiltonian must be such that that there exists a tensor-product structure (TPS) $\cal{H}\sim \cal{H}_C\otimes \cal{H}_R$, where the first subsystem represents the clock and the second the rest of the universe, \cite{PER,PAWO}, such that this crucial {\sl non-interacting property} holds: $$H=H_C\otimes{\mathfrak I}+{\mathfrak{I}}\otimes H_R$$ where $\mathfrak{I}$ denotes the unit operator on each subspace. 

In classical physics, the \qq{measurement of time} is always performed relative to some dynamical variable (e.g. a pointer on a clock dial). In quantum theory, a similar logic is valid \cite{PER}. For the ideal clock, the observable to choose as indicator is the {\sl conjugate} observable $T_C$ to the clock Hamiltonian, $[{H}_C,T_C]=i$, with  $T_C\ket{t}=t\ket{t}$, where the values $t$ form a continuum, which represent the values to be read on the hands of the clock. Once more, note that $T_C$ is {\sl not} a time-operator. It is an observable of the clock subsystem. 

That clocks are possible in reality means that the behaviour of the ideal clock can be approximated to an arbitrarily high accuracy: as pointed out in \cite{DEU}, the ideal clock can be approximated by systems with an observable $T$ that has a discrete spectrum of $n$ values $t_n$, where there is no limit to how well the sequence of $t_n$ can approximate the real line.  In this paper we shall confine our attention to the ideal case, for simplicity of exposition. 

{\bf Entanglement.} The third sufficient condition for the P-W construction to hold is that the clock and the rest of the universe are {\sl entangled}: as it will become clear in a moment, this is the feature that allows the appearance of dynamical evolution on the rest to be recovered out of no evolution at all at the level of the universe. Formally, this means that the state of the universe $\ket{\psi}$ must have this form:

\begin{equation}
\ket{\psi}=\sum_t\alpha_t\ket{t}\ket{\phi_t}\;\label{dec}
\end{equation}

for some appropriate $\ket{\phi_t}$ defined on the rest, with two or more of the $\alpha_t$ being different from zero. In practice, as we shall see, for this to produce a realistic dynamics, $\alpha_t\neq 0$ for a sufficiently large number of $t$'s. This is because all that happens in the rest is given once and for all in the state $\ket{\psi}$. By taking one of the clock eigenstates $\ket{0}$ as the initial time, whereby $\ket{t}=\exp{(-i H_C t)}\ket{0}$, the story of the rest of the universe is a sequence of events encoded in the various $\ket{\phi_0}, \ket{\phi_1}, ...,\ket{\phi_t}$.  

Note that the rest and the clock must {\sl not} be in an eigenstate of their local Hamiltonians, otherwise the dynamics is trivial. In the basis $\ket{\epsilon_n}\ket{E_n}$ defined by the local hamiltonians $H_C$ and $H_R$, the  universe state is therefore $\ket{\psi}=\sum_{m,n}\psi_{m,n} \ket{\epsilon_m}\ket{E_n}$ where  $H\ket{\epsilon_m}\ket{E_n}=0\;\; \forall n,m $. An elementary example will clarify this point. Consider a universe made of two qubits only, with Hamiltonian $H=\sigma_z\otimes{\mathfrak I}+{\mathfrak{I}}\otimes \sigma_z$, where $\sigma_z$ represents the z-component of the spinor $\left (\sigma_x, \sigma_y, \sigma_z\right)$,  $[\sigma_i, \sigma_j]=2\epsilon_{i,j,k}\sigma_k$ and $\sigma_i\sigma_j=2\delta_{i,j}$. The clock observable can be  $\sigma_x$, so that in the clock basis $\ket{+},\ket{-}$ the state of the universe can be written as $\ket{\psi}=\frac{1}{\sqrt{2}}\left (\ket{+-}+\ket{-+}\right)$. As required, the Hamiltonian of the clock generates the shift on the two clock \qq{hands},  $\exp(-i\sigma_z\frac{\pi}{2})\ket{+}=\ket{-}$.  In the energy basis  (the basis of eigenvectors of $\sigma_z$) the state of the universe is $\ket{\psi}=\frac{1}{\sqrt{2}}\left (\ket{01}+\ket{10}\right)$. 

Therefore for this construction to be compatible with a realistic dynamics there must be a high degree of degeneracy in the Hamiltonian $0$-eigenspace.

\medskip 

If the above conditions are satisfied, the evolution without evolution can be reconstructed as follows. The state of the rest of the universe when the clock reads $t$ is the Everett relative state, \cite{EVE}, defined as:

\begin{equation}
\rho_t=\frac{{\rm Tr_c}\{P_t^{(c)}\rho\}}{{\rm Tr}\{P_t^{(c)}\rho\}}=\ket{\phi_t}\bra{\phi_t}.
\end{equation}

Note that the projector in the definition of relative state has nothing to do with measurement and does not require one to be performed on the clock: rather, the relative states are a 1-parameter family $\ket{\phi_t}$ of  states, labelled by $t$, each describing the state of the rest with respect to the clock ‘given that’ the latter is in the state $\ket{t}$. By using the constraint $\eqref{tot}$, the special, {\sl non-interacting} form of $H$, and the fact that $[H_C,T_C]=i$, one obtains that the relative state of the rest evolves according to the Schr\"odinger equation with respect to the parameter $t$:

\begin{equation}
\frac{\partial \rho_t}{\partial t}= i[\rho_t, H_R]\;.\label{SC}
\end{equation}

Thus, the logic that \qq{time can be said to exist if there is a description of the physical world such that the Schr\"odinger  (or Heiseberg) equation holds on a subsystem of the universe}, seems to be applicable to quantum theory. The parameter $t$ is to be interpreted as time, and the evolution of the rest of the universe has been recovered out of no evolution at all. 

Assuming that the eigenstates of the clock have the form $\ket{t}=\exp(-iH_Ct)\ket{0}$ may seem too strong a constraint: together with the fact that the clock and the rest are entangled, that constraint directly implies that the evolution on the rest has to have the same exponential form leading to the Schr\"odinger equation. However, the main point of the PW approach is to show that that there exists at least one such choice. This is a rather remarkable property of unitary quantum theory, as it implies that it is consistent with there being the appearance of dynamics in a subpart of the universe, even when the whole universe is at rest. 

Note also that this construction is compatible with a time-dependent Hamiltonian arising on a {\sl subsystem} of the rest, just like in ordinary quantum mechanics. The time-dependent Hamiltonian for the subsystem only is an approximate description, generated by the interactions (encoded in the time-independent hamiltonian $H_R$) between the subsystem and the environment, in the approximation where the environment can be treated semi-classically (see, e.g. \cite{JAB}).

\section{There is no ambiguity}

A problem seems to arise in the PW logic. Quantum theory provides infinitely many inequivalent ways of partitioning the total Hilbert space of the universe into a tensor-product structure (TPS); as a consequence, there would seem to be several choices of the clock by which unitary evolution can arise on the rest of the universe. If true, this would mean that given the same overall state $\ket{\psi}$ describing the universe, the PW approach leads to completely different dynamics on the rest of the universe. This is the so-called \qq{clock ambiguity}, \cite{ALB}. We are about to show that this ambiguity does not in fact arise: having fixed the total Hamiltonian and the overall state $\ket{\psi}$ of the universe, if there is \textit{one} tensor-product structure - i.e., one partition of the universe into a good clock and the rest - leading to a unitary evolution generated by a {\sl time-independent} Hamiltonian for the relative state, then it must be unique. The crucial property will be that the clock is not any old subsystem of the universe, but it must be, in the ideal case, a {\sl non-interacting} one.
 
Let us first summarise the clock ambiguity problem. By choosing a suitable orthonormal basis $|{k}\rangle $ in the overall Hilbert space ${\cal H}$, one can write: $$\ket{\psi} = \sum_k 
\alpha_k |{k}\rangle\;,$$ where $\ket{k}\iff \ket{t}_C\ket{\phi_t}_R$ in a given tensor product structure ${\cal H}\sim {\cal H_C}\otimes {\cal H_R} $. 

The clock ambiguity is 
thus expressed: consider a different state of the universe, such as $$|\tilde 
{\psi}\rangle = \sum_k \beta_k |{k}\rangle\;.$$ 

There is of course a unitary operator $W$ such that $|\tilde {\psi}\rangle = W|{\psi}\rangle$. Hence \begin{equation}|{\psi}\rangle =\sum_k \beta_k 
W^{\dagger}|{k}\rangle =\sum_k \beta_k |\tilde{k}\rangle\label{AMB}\end{equation} where we have defined $\ket{\tilde k}=W^{\dagger}|{k}\rangle$.

Now, it is possible to choose a {\sl different} bi-partite tensor-product structure whereby: $ |\tilde{k}\rangle = |t\rangle_{\tilde C}|\phi_t\rangle_{\tilde R}$. The clock ambiguity is that there are countless such choices, and 
they would each seem give rise to very different description of the evolution of the rest. In one, the rest would appear to evolve according to the sequence of relative states: $\ket{\phi_0}_R$,$\ket{\phi_1}_R, ...., \ket{\phi_t}_R$; in the other, it would go through the sequence of {\sl different} relative states: $\ket{\phi_0}_{\tilde R }, \ket{\phi_1}_{\tilde R },....,  \ket{\phi_t}_{\tilde R }$. 

In fact, the clock ambiguity does not arise, because the PW model has additional constraints. In short, a clock is a special subsystem of the universe, which must not interact with the rest, in the ideal case. So, let us assume that there exists a tensor-product structure $\cal{H}\sim \cal{H}_C\otimes \cal{H}_R$ where the clock and the rest are non-interacting: $H=H_C\otimes{\mathfrak I}+{\mathfrak{I}}\otimes H_R$ -- whereby, applying the PW argument, the relative state $\ket{\phi_t}_R$ of the rest evolves according to a unitary evolution generated by $\exp{(-iH_Rt)}$. 

Formally, a tensor product structure is a unitary mapping $U$ whose elements  $U_{a,b}^{k}$ have the property that, for any state $\ket{k}\in{\cal H}$ and some basis states $\ket{a}_C\ket{b}_R$: $$ |{k}\rangle = \sum_{a,b}U_{a,b}^{k}|a\rangle_C|b\rangle_R.$$

Two tensor product structures are {\sl equivalent} if and only if their elements $U_{i,j}^{k}$, $\tilde U_{a,b}^{k}$ are related by {\sl local} unitaries $P$, $Q$: $$ U_{i,j}^{k}= \sum_{a,b}P_{i}^{a}Q_{j}^{b}\tilde U_{a,b}^{k}\;.$$ 

Hence, the case where the new TPS is equivalent to the original one corresponds to $W=P\otimes Q$ in \eqref{AMB}, i.e., to choosing a different clock observable $P^{\dagger}T_CP$ from the optimal one $T_C$ (the conjugate observable to $H_C$). Therefore this case need not concern us any further, as it simply consists of choosing a poorer clock.

The case where the new TPS is not equivalent requires a little more explanation. In this case, the unitary $W$ in equation \eqref{AMB} has the form: $W=\exp\{-i\left(W_C+W_R+W_{CR}\right)\}$, for some Hermitean  operators $W_C$, $W_R$ $W_{CR}$, where $W_{CR}$ operates as an interaction term between the two subsystems $C$ and $R$ of the original TPS. For two qubits, the most general form is: $$W_{CR}=\sum_{\alpha,\beta\in\{x,y,z\}}w_{\alpha,\beta}\;\sigma_{\alpha}\otimes\sigma_{\beta}\;$$ for real coefficients $w_{\alpha, \beta}$. The cases where $[H,W]=0$ or $[H,W_{CR}]=0$ also need not concern us any further, because in both cases $W$ would have a trivial, local action on $\ket{\psi}$.  

The remaining case can be addressed as follows. $H$ is the sum of two non-interacting terms for $C$ and $R$ in the tensor-product structure defined by $U_{i,j}^{k}$. Therefore, in {\sl any} tensor product structure $\tilde U_{a,b}^{k}$ obtained via $W$ acting on  the TPS defined by $U_{i,j}^{k}$, $H$ will have an interaction term between the new clock  $\tilde C$ and the new rest $\tilde R$ :
 
 \begin{equation}
 H=H_{\tilde C}\otimes{\mathfrak{I}}+{\mathfrak{I}}\otimes H_{\tilde R}+V_{\tilde C}\otimes V_{\tilde R}\;
 \end{equation} 
because the transformation to the new TPS is generated by a non-local unitary transformation. As a consequence, in the new, non-equivalent, tensor-product structure, the evolution of the relative state as a function of the labels of the eigenstates of observable $T_{\tilde C}$ of the clock will \textit{not} be a unitary evolution generated by a \textit{time-independent} Hamiltonian. As pointed out in \cite{PAGE} it will have the form:

\begin{equation}
\frac{\partial \rho_t}{\partial t}= i[\rho_t, H_{\tilde R}]\;+ {\rm terms\; depending\; on\; t}.\label{rot}
\end{equation}

Hence, given $H$ and $\ket{\psi}$, if there is {\sl one} tensor product structure in which the clock is ideal (no interactions) {\sl and} a Schr\"odinger-type unitary evolution (generated by the time-independent hamiltonian $H_R$)  arises on the relative state of the rest with respect to the labels $t$, then the TPS must be unique. In all other non-equivalent tensor product structures, although it is possible to write the overall state as 
$|{\psi}\rangle =\sum_t \beta_t \ket{t}_{\tilde C}\ket{\phi_t}_{\tilde R}$, it must be $\ket{\phi_t}_{\tilde R}\neq \exp(-iH_{\tilde R}t)\ket{\phi_0}_{\tilde R}$, because the eq. \eqref{rot} holds instead of \eqref{SC} -- due to the interaction terms between the clock $\tilde C$ and the rest $\tilde R$. Thus, there is no clock ambiguity, as promised.

We conclude that in unitary quantum theory it is not ambiguous to apply the same logic as in the classical time-less approaches: the clock and the clock observable are to be chosen so that a Schr\"odinger-type  dynamics arises on the rest of the universe, generated by a time-independent Hamiltonian. There can be only one such choice, for a given total Hamiltonian $H$ and a given total state of the universe.

{\bf The appearance of the flow of time.}  It is worth pointing out that there is no flow of time in the PW picture. The PW approach shows that the Schr\"odinger equation generated by $H_R$ holds for the rest of the universe with respect to the labels $\{t\}$ of the eigenstates of a particular clock observable $T_C$, conjugate to $H_C$, $\ket{t}=\exp(-iH_Ct)\ket{0}$. But the {\sl flow of time} has to emerge as a result of there being subsystems of the rest of the universe that can perform measurements and store their results, thus constructing a {\sl history}.  More specifically, let us consider a model where the rest is partitioned in three subparts, the \qq{observer} (which for simplicity we assume to be made of a memory only), the \qq{observed} and a sequence of ancillas. As mentioned in section 3, by treating the ancillas semiclassically, it is possible to describe the observed and the observer as undergoing an effective evolution generated by a time-dependent Hamiltonian, in turn generated by the interactions with the ancillas as prescribed by $H_R$ -- where time here is the label $t$ of the eigenstates of the clock. 

Let us suppose that effective evolution corresponds to a sequence of gates occurring on the observed and to the observer performing measurements on the observed to record what has happened.  Specifically, suppose that the observer's memory starts in a blank state $\ket{b}^{\otimes N}$ at $t=0$ and the observed is in a state $ \ket{A_1}$ where $\ket{A_i}$, for $i=1...N$, is an eigenstate of some observable $A$. Suppose that the observer and the observed evolve under the effective time-dependent hamiltonian as follows: at time $t=1$ the observer measures the observable  $A$, so that its state changes to $\ket{\text{Saw}A_1}\ket{b}^{\otimes N-1}\ket{A_1}$; then a local permutation happens on the observed, so that the state changes to  $\ket{\text{Saw}A_1}\ket{b}^{\otimes N-1}\ket{A_2}$; then the observer measures $A$ again, so that the state is now $\ket{\text{Saw}A_1}\ket{\text{Saw}A_2}\ket{b}^{\otimes N-2}\ket{A_2}$ and so on, until the observer ends up recording a sequence of events $A_1, A_2, ...,A_N$ as prescribed by $H_R$. All these events, here described as sequential, are encoded statically in the overall state of the universe $\ket{\psi}$. 

The beauty of the PW approach is that it is fully internally consistent. The observer cannot empirically tell the difference between a situation where the sequence of events it observes is really generated by the Hamiltonian $H_R$ on the rest and all the other possibilities, which include cases where the universe can be manipulated by some external entity. This is a general feature of any other picture where constructing a history of events is possible in the sense above, e.g. static pictures such as the Block Universe in general relativity. Imagine, for example, that an entity existing outside of the PW universe were able to decide which element in the PW wavefunction constitutes the \qq{now}, as defined with respect to some external coordinate time --  a sort of \qq{meta-time} existing outside of the PW universe. The entity is able to point at any one element $t_n$ declaring it the now. Then (again in the meta-time picture) it could point at another element $t_m$ as the next now. And so on. This corresponds to picking a different generator of the clock states than $\ket{t}=\exp(-iH_Ct)\ket{0}$, corresponding to a different dynamical law from the $H_R$-generated Schr\"dinger equation. The entity can choose any order it likes for the labels $t$; not necessarily the one corresponding to the sequence of events outlined above. What if the entity decides to point first at a state which in the original labeling appears \qq{later}, and then at an \qq{earlier} one? This may seem to make time flow backwards even from the point of view of the observer. But that is not the case.

The observer, from his own perspective, does not notice anything different, as his state only contains information about the \qq{previous times}. For example, the observer could never see any gaps if the entity decides to jump, say, from time $t=1$ to time $t=100$, because in the observer state corresponding to $t=100$ there will be recollections of all events $E_2...E_{100}$, from $t=2$ up to $t=100$. As far as the observer is concerned he perceives himself as coming to time $t=100$ directly from $t=99$. Therefore any experiment made by the observer would lead him to conclude that what he observes is consistent with the same Schr\"odinger equation as that corresponding to there being no jumps at all (in the meta-time), generated by the hamiltonian $H_R$. In other words, in the PW approach just like in any other dynamical theory, the existence of a meta-time is completely irrelevant from the observer's perspective, as it would not have any empirical consequences.

{\bf The arrow of time.}  In the PW picture there is no arrow of time, just like in unitary quantum theory: the PW approach gives rise to a time-reversal symmetric dynamical law. Thus, the arrow of time has to be imposed by a separate postulate, requiring that under that dynamical law a given monotone always increases (or decreases). Namely, suppose again that the rest consists of two subsystems, \qq{the observer} and \qq{the observed}. For simplicity, let us approximate the \qq{observer} as simply consisting of a memory needed to keep track of what happens to the observed. The arrow of time can now be specified by the increase in entanglement between the observer and the observed, by selecting a measure of entanglement, and requiring that entanglement never decreases on the rest under the evolution generated by $H_R$. In other words, early times correspond to no entanglement between the observer and the observed and later times to more and more entanglement (as the observer learns more and more about the system). Since the relative states of the rest are pure states, when considering a bi-partition there is a unique measure, which consists of the relative entropy \cite{VE}. For a discussion of some explicit models, see \cite{PAGE}.  The only ambiguities that might arise in this context are due to: 1) the possibility of picking different partitions into subsystems; and 2) the possibility of having a partition into n-subsystems, where $n\geq 3$: for, in this case, there is not a unique measure. However, these ambiguities are the same as those related to which coarse-graining to adopt in the usual statistical-mechanical picture.  Hence this is no more problematic than any other coarse-graining approaches to irreversibility in statistical-mechanics.

\section{Conclusions}

We have shown that the PW timeless approach to time in quantum theory has no ambiguities, thus vindicating it as a viable proposal for the emergence of time in a time-less universe described by the unitary quantum theory. The non-interacting property of the clock is crucial to establish that result: a good clock is not just a system with a large set of orthogonal states, such as good memory; it must be non-interacting while it is used as a clock. 

We have also updated the model so that it becomes possible to apply to to more general theories, including the successor of quantum theory. One possible development is to investigate under what conditions the PW logic could apply to different theories than quantum theory (e.g.,  generalised probabilistic theories \cite{BAR1} or constructor theory's super-information theories \cite{MA}). The challenge there is to understand what relative states would be, as well as in what form the clock ambiguity might appear. Another interesting application could be to recast this model in terms of pseudo-density matrices, \cite{PSEUDO} where time and space are treated in a unified framework.  Finally, it is worth speculating about how the PW approach might provide observable consequences, when combined with cosmological models. For example, in the context of an expanding universe, one might use as clock observable the radius of the universe, whereby $\ket{\psi}=\sum_n\alpha_n\ket{t_n}\ket{\phi_n}\;$, where $t_n$ is the radius of the universe. In such models, \cite{REF}, $\langle t_n|t_m\rangle\sim \exp(-\gamma(t_n-t_m))$, where $\gamma$ is some parameter that can be fixed according to the particular cosmological model -- which means that different states of the clock get more and more distinguishable the more they are separated in \qq{time}. When applied in this scenario, the PW construction would lead to the conclusion that the relative state of the rest of the universe is no longer pure, but it is a mixed state -- this is because the operators $P_t=\ket{t}\bra{t}$ involved in constructing the relative state are no longer orthogonal projectors. This fact might have observable consequences even at the present epoch, according to which particular cosmological model one chooses, provided that the accuracy for measuring time is high enough. We leave investigating all that to future work.

\section{Acknowledgements}
The authors thank David Deutsch for fruitful discussions and comments; Mario Rasetti for helpful suggestions. CM's research was made possible through the support of a grant from the Templeton World Charity Foundation. The opinions expressed in this publication are those of the authors and do not necessarily reflect the views of Templeton World Charity Foundation. VV thanks the Oxford Martin School, Wolfson College and the University of Oxford, the Leverhulme Trust (UK), the John Templeton Foundation, the EU Collaborative Project TherMiQ (Grant Agreement 618074), the COST Action MP1209, the EPSRC (UK) and the Ministry of Manpower (Singapore). This research is also supported by the National Research Foundation, Prime Minister’s Office, Singapore, under its Competitive Research Programme (CRP Award No. NRF- CRP14-2014-02) and administered by Centre for Quantum Technologies, National University of Singapore.

 \end{document}